\documentclass[aps,prd,fleqn,superscriptaddress]{revtex4}
\usepackage{graphicx,color,natbib}
\usepackage{amsmath,amssymb,amsfonts}
\newcommand{\bse}{\begin{subequations}}
\newcommand{\ese}{\end{subequations}}
\newcommand{\be}{\begin{equation}}
\newcommand{\ee}{\end{equation}}
\newcommand{\bea}{\begin{eqnarray}}
\newcommand{\eea}{\end{eqnarray}}
\newcommand{\ba}{\begin{array}}
\newcommand{\ea}{\end{array}}
\newcommand{\h}{\frac{1}{2}}

\def\FF{{\mathcal{F}}}
\def\HH{{\mathcal{H}}}
\def\B{{\mathcal{B}}}

\input amssym.def
\input amssym.tex

\usepackage[colorlinks=true, linkcolor=blue, bookmarks=true]{hyperref}
\begin{document}
\title{Holographic Entanglement Entropy Decomposition in an Anisotropic Gauge Theory }
%
\author{M. Rahimi}
\email{{\rm{me}}$_{}$ rahimi@sbu.ac.ir}
\affiliation{Department of Physics, Shahid Beheshti University G.C., Evin, Tehran 19839, Iran}
\author{M. Ali-Akbari}
\email{{\rm{m}}$_{}$ aliakbari@sbu.ac.ir}
\affiliation{Department of Physics, Shahid Beheshti University G.C., Evin, Tehran 19839, Iran}
\begin{abstract}
We study holographic entanglement entropy in spatially anisotropic field theory. We observe that for the background we consider in this paper, to a good approximation, the holographic entanglement entropy can be decomposed into two terms. One of them is the entanglement entropy of the isotropic field theory at fixed temperature and the other term is only a function of anisotropy parameter. Moreover, for large enough values of anisotropy parameter, our numerical results indicate that the entanglement entropy in the perpendicular direction to anisotropic direction is greater than the parallel case.
\end{abstract}
\maketitle
\tableofcontents
%
\section{Introduction}
During two last decades, gauge-gravity duality, as a new framework, received a lot of interest to explain the physics of strongly coupled field theories. Since standard methods such as perturbative expansion are not applicable due to large coupling constant, gauge-gravity duality plays an important role in this area of physics. This duality states that a strongly coupled gauge field theory in $d$-dimensional space-time corresponds to a classical gravity in $d+1$-dimensional background \cite{Maldacena}. According to the gauge-gravity duality, the filed theory is living on the boundary of the background. All parameters, fields and processes in the gauge theory are then translated into appropriate equivalent on the gravity side. Up to now, this duality has been frequently applied and reveals a lot of valuable information about strongly coupled field theories. For a good review with an extensive list of references see \cite{solana}.
For describing various properties of strongly coupled gauge theories, this duality usually proposes applicable and simple prescriptions in gravity theory. In other words, although explaining of different properties may not
seem simple from the field theory perspective, its gravitational description is generally more tractable. As an example, in order to calculate the entanglement entropy in the field theories with holographic dual there is a very simple prescription firstly introduced in \cite{Ryu:2006ef}. The entanglement entropy is analytically calculated in two-dimensional conformal field theory and its generalization to higher dimensions is not obvious. However, as we will see in the next section, its gravitational dual is simple and can be generalized to arbitrary dimension. This conjectured formula does indeed satisfy many non-trivial relations known in quantum information theory.
In this paper, using the holographic description of the entanglement entropy, we would like to investigate the effect of the anisotropy parameter $a$ on the entanglement entropy in an anisotropic thermal gauge field theory. Here "anisotropic" refers to spatially anisotropic systems or equivalently the pressure in one direction, say $z$, is different from the others. For this reason, we will work with the anisotropic background introduced in \cite{Mateos:2011tv}. Then we consider the entangling length in the anisotropic direction $z$, parallel case, or in the transverse directions, i.e. perpendicular to $z$, perpendicular case. According to our numerical outcomes, the main findings can be summarized as follows:
\begin{itemize}
\item At fixed temperature $T$, the entanglement entropy for the perpendicular case is greater than parallel case, that is $S_A^\bot>S_A^\parallel$ for $a>2T$. The same behavior is also true at fixed entropy density $s$ for $a>2s^{1/3}$. It is important to notice that this behavior does not persist in high temperature limit, i.e. $T\gg a$.
\item At fixed temperature, for both parallel and perpendicular cases, to a good approximation, the entanglement entropy can be decomposed into two terms: one of them is the entanglement entropy of the isotropic field theory at finite temperature and the other one is independent of temperature and depends only on the anisotropy parameter. This statement is also true when the entropy density is kept fixed. Interestingly, this decomposition can be done in the high temperature limit, too.
\item In the desired range of anisotropy parameter the term which depends only on the anisotropy parameter is proportional to $k_1 a^4-k_2 a^2 (-n_1 a^2+n_2 a^4)$ for perpendicular (parallel) case where $k_1, k_2, n_1$ and $n_2$ are positive numbers. Independent of temperature and entropy density which are kept fixed, the coefficients $k_1$ and $k_2$ ($n_1$ and $n_2$) are the same, interestingly.
\item Making the anisotropy parameter larger, the deviation of anisotropic entanglement entropy form the isotropic one becomes greater when the temperature or entropy density is kept fixed.
\item We checked and believed that all results reported in this paper exist for any acceptable entangling length.
\end{itemize}
\section{Review on (Holographic) Entanglement Entropy}
\begin{figure}[ht]
\begin{center}
\includegraphics[width=70mm]{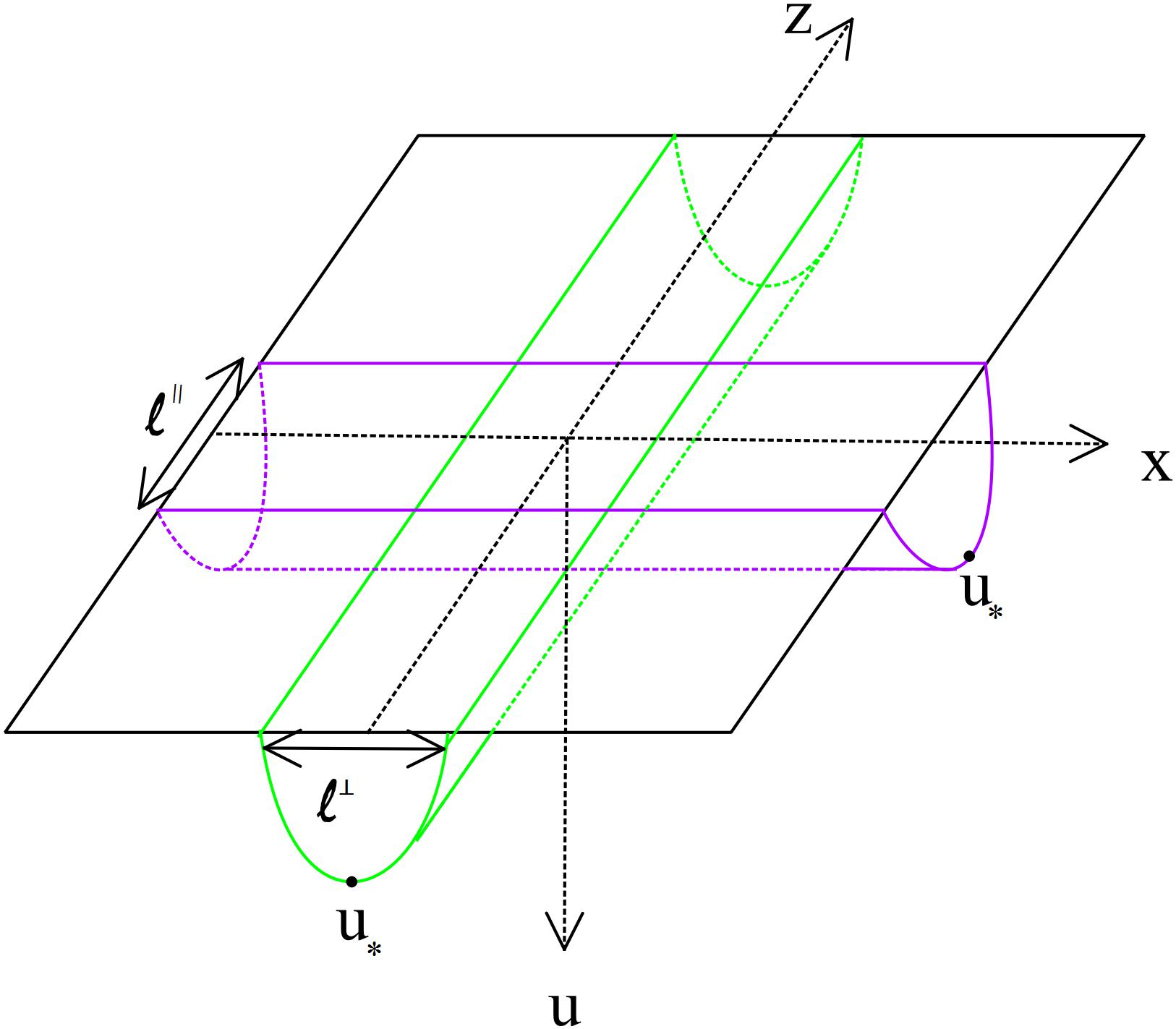}
\caption{Minimal surfaces in the anisotropic background.}\label{fff}
\end{center}
\end{figure}

A well-known non-local observable in the quantum field theory is entanglement entropy \cite{thesis}. Consider a quantum field theory whose state is described by the density matrix $\rho$. The entanglement entropy of a spatial subregion $A$, with complement $B$, denotes how much entanglement exists between $A$ and $B$ and it is given by
\be %
S_A=-Tr(\rho_A \log\rho_A),
\ee %
where $\rho_A=Tr_{B}(\rho)$ is reduced matrix and obtained by tracing over the degrees of freedom in the region $B$. The entanglement entropy of $A$ shows the amount of information lost when an observer is limited to the subregion $A$. Moreover, it has a main divergency which is proportional to area of the subregion $A$ for space-time dimension greater than two. Although calculating the entanglement entropy is normally difficult, it has hopefully a famous and simple description in the context of the gauge-gravity duality. In fact,
Ryo and Takayanagi firstly proposed in \cite{Ryu:2006ef} that the entanglement entropy can be computed from
\be \label{rt}%
S_A=\frac{\rm{Area}(\gamma_A)}{4G_N^{d+2}},
\ee %
where $G_N^{d+2}$ is the $d$-dimensional Newton constant. $\gamma_A$ is a codimension-2 minimal surface whose boundary $\partial\gamma_A$ coincides with the boundary of the subregion $A$ on the boundary of the bulk where the quantum field theory lives, i.e. $\partial\gamma_A=\partial A$. This proposal received a lot of interest during the last decade and passed several non-trivial checks known in the quantum field theory. For more details, we refer the reader to \cite{Headrick:2007km}.
In order to calculate the holographic entanglement entropy, we start with a $d+2$-dimensional background
\be\label{metric1}
ds^2=-f_1(u)dt^2+f_2(u) du^2+f_3(u)dz^2+f_4(u)d\vec{x}^2,
\ee
where $u$ is radial direction and the boundary is located at $u=0$. $f_1,\dots ,f_4$ are arbitrary functions and depend only on the radial direction. $(t,z, \vec{x}=x_1,\dots, x_{d-1})$ represent $d+1$-dimensional boundary coordinates where the field theory lives. We also demand that the background approaches $AdS_{d+2}$ with radius one asymptotically. As it is clearly seen, since $f_3\neq f_4$ generally, the above background is anisotropic along the gauge theory directions and hereafter we call $z$ the \textit{anisotropic coordinate}. $f_1$ and $f_2$ are blacking factors that vanish at the position of the horizon, i.e. $f_1(u=u_h)=f_2^{-1}(u=u_h)=0$ and therefore the above metric is a black hole solution. According to the gauge-gravity duality, the Hawking temperature of the black hole is identified with the temperature of the field theory.
We now consider the simplest shape for the boundary entangling region $A$ which is a rectangular shape with one dimension of length $l$ in $z$ (parallel to the anisotropic coordinate) or $x\in(x_1,\dots, x_{d-1})$ (perpendicular to the anisotropic coordinate) direction at a constant time slice and all other coordinates have infinite width, see figure \ref{fff}. Note that since there is a rotational symmetry in $\vec{x}$ directions, there is no difference among them.
The recipe (\ref{rt}) states that the entanglement entropy between the regions $A$ and its complement $B$ is proportional to the minimal surface $\gamma_A$ and it can be then obtained from
\be\label{area} %
S_A=\frac{1}{4G_N^{d+2}}\int d^{d}\sigma \sqrt{\det g_{ab}},
\ee %
where $g_{ab}$ is induced metric on $ \gamma_A $ and is defined as $g_{ab}=G_{MN} \partial_a X^M \partial_b X^N$. $ X^M$s are bulk coordinates introduced in \eqref{metric1} and $\sigma^a$s denote coordinates on the $\gamma_A$ and we thus have $X^M(\sigma^a)$. In the perpendicular case, using static gauge, i.e. $(\sigma^1, \dots, \sigma^{d}) \equiv (z, x^1, \dots, x^{d-1})$, and due to translational symmetry in the all directions in the background except $x\equiv x^{d-1}$, the shape of the minimal surface corresponding to the rectangular stripe is described by $u(x)$, see figure \eqref{fff}. Similarly, for the parallel case the minimal surface can be described by $u(z)$. In the following, using the gauge-gravity duality, we proceed our calculations to obtain the entanglement entropy for these two cases.
\begin{itemize}
\item \textbf{Perpendicular case:}
In this case we have $u(x)$ and after simple calculation the holographic entanglement entropy \eqref{area} leads to
\begin{equation}\label{ee1}
S_A^\bot = \frac{{{V_{d - 1}}}}{{4{G_N^{d+2}}}}\int_{-\frac{l}{2}}^{\frac{l}{2}} d x\sqrt {{f_3}(u)f_4^{d - 1}(u) + {f_2}(u)f_4^{d - 2}(u){f_3}(u){u^{\prime 2}}},
\end{equation}
where $u'=\frac{du}{dx}$ and $V_{d-1}$ is the volume of the isotropic transverse coordinates. Since the above equation does not depend explicitly on the $x$, a constant of motion can be easily found as
\be
\frac{{{f_3(u)}f_4^{d - 1}\left( u \right)}}{{\sqrt {{f_3}(u)f_4^{d - 1}(u) + {f_2}(u)f_4^{d - 2}(u){f_3}(u){u^{\prime 2}}} }} = {f_3}({u_*})f_4^{d - 1}({u_*}),
\ee
where $u_*$ is the turning point at which $u'(x)=0$ and is located at $x=0$ by symmetry. It then turns out
\be\label{uprim1}
{u^\prime } = \left( \frac{f_4}{f_2}\bigg(\frac{f_3 f_4^{d - 1}}{f_{3*}f_{4*}^{d - 1}} - 1 \bigg) \right)^{\frac{1}{2}},
\ee
where $f_{3*}=f_3(u_*)$ and so on. From the above equation one gets
\be
l^\bot = 2\int_0^{u_*} du \left( \frac{f_4}{f_2}\bigg(\frac{f_3 f_4^{d - 1}}{f_{3*}f_{4*}^{d - 1}} - 1 \bigg) \right)^{-\frac{1}{2}}.
\ee
Finally, by substituting (\ref{uprim1}) into \eqref{ee1}, the holographic entanglement can be obtained
\be\label{eev}
S_A^\bot = \frac{V_{d - 1}}{2{G_N^{d+2}}}\int_0^{u_*} du {\frac{\sqrt {f_2} f_4^{d - \frac{3}{2}}{f_3}}{{\sqrt {{f_3}f_4^{d - 1} - {f_{3*}}f_{4*}^{d - 1}} }}}.
\ee
\item \textbf{Parallel case:}
In this case, as it was already explained, we have $u(z)$. Similar to the perpendicular case, the entanglement entropy and entangling length $l^\parallel$ can be found
\bse\begin{align}\label{eep}
S_A^{\parallel}&=\frac{V_{d-1}}{2G_N^{d+2}}\int_{0}^{u_*} du \frac{\sqrt{f_2 f_3}f_4^{d-1}}{\sqrt{f_3 f_4^{d-1}-f_{3*}f_{4*}^{d-1}
}},\\
l^\parallel &= 2\int_0^{u_*} du \left( \frac{f_3}{f_2}\bigg(\frac{f_3 f_4^{d - 1}}{f_{3*}f_{4*}^{d - 1}} - 1 \bigg) \right)^{-\frac{1}{2}}.
\end{align}\ese
\end{itemize}
A specific solution of \eqref{metric1} is isotropic black hole background which is given by
\be %
u^2 f_1(u)=u^{-2} f_2^{-1}(u)=1-\frac{u^4}{u_h^4},\ \ \ \ f_{3}(u)=f_4(u)=\frac{1}{u^2}.
\ee %
Obviously, for the above solution we get $S_A^0=S_A^\bot=S_A^\parallel$. As it is well-known, the entanglement entropies $S_A^\bot$ and $S_A^\parallel$ are divergent and this divergency is proportional to the area of subregion $A$. In order to find finite entanglement entropies, we define the following functions
\bse\begin{align}
\label{a1} \Delta S^\bot&=\frac{G_N^{d+2}}{V_{d-1}}(S_A^\bot-S_A^0),\\
\Delta S^\parallel&=\frac{G_N^{d+2}}{V_{d-1}}(S_A^\parallel-S_A^0).
\end{align}\ese
where $\Delta S^\bot$ and $\Delta S^\parallel$ are now finite. We emphasize here that these two quantities represent the difference between entanglement entropies in anisotropic and isotropic background which will be computed at fixed temperature or fixed entropy density. More specifically, they provide noteworthy information about the effect of anisotropy parameter on the entanglement entropies, as we will see in next section.
\section{Numerical Results}
The background we are interested in is an anisotropic solution of IIB string theory introduced in \cite{Mateos:2011tv} and different aspects of this solution have been illustrated in the literature, for instance see \cite{Ali-Akbari:2014xea}. Regarding \eqref{metric1}, the anisotropic metric with $d=3$ is
\begin{equation}\label{metrica}
{f_1}(u) = {\cal F}{\cal B}{u^{ - 2}},\ \ \ \ {f_2}(u) = {{\cal F}^{ - 1}}{u^{ - 2}},\ \ \ \
{f_3}(u) = {\cal H}{u^{ - 2}},\ \ \ \ {f_4}(u) = {u^{ - 2}}.
\end{equation}
The functions $\HH$, $\FF$ and $\B$ depend only on the radial direction.
In terms of the Dilaton field, they are %
\bse\label{three}\begin{align} %
\HH&=e^{-\phi},\\
\label{ff} \FF&= \frac{e^{-\frac{1}{2}\phi}\left[ a^2 e^{\frac{7}{2}\phi}(4u+u^2\phi')+16\phi'\right]}{4(\phi'+u\phi'')}\,,
\\
\frac{\B'}{\B}&=\frac{1}{24+10
u\phi'}\left(24\phi'-9u\phi'^2+20u\phi''\right)\,,
\end{align}\ese %
where the Dilaton field satisfies the following third-order equation %
\begin{eqnarray}\label{eomphi}%
&&\frac{256 \phi ' \phi ''-16 \phi '^3
\left(7 u \phi '+32\right)}{u \, a ^2 e^{\frac{7 \phi }{2}}
\left(u \phi '+4\right)+16 \phi '}
+\frac{\phi ' }{u \left(5 u \phi '+12\right) \left(u \phi
''+\phi '\right)}\nonumber
\times \Big[13 u^3 \phi '^4+8 u \left(11
u^2 \phi ''^2-60\phi''-12 u \phi ''' \right)\nonumber \\
&&+ u^2 \phi '^3
\left(13 u^2 \phi ''+96\right)
+2 u \phi '^2 \left(-5 u^3 \phi
'''+53 u^2 \phi ''+36\right)\nonumber
+ \phi ' \left(30 u^4 \phi
''^2-64 u^3 \phi '''-288+32 u^2 \phi
''\right) \Big]=0 \,.
\end{eqnarray}%
Note that the solution also contains a self dual five-form field.
The horizon is located at $u=u_h$ where $\FF(u_h)=0$. The Hawking temperature and entropy density per unit volume are given by \cite{Mateos:2011tv} %
\be\label{entro}\begin{split} %
T&=-\frac{1}{4\pi}\FF'(u_h)\sqrt{\B(u_h)},\cr
s&=\frac{\pi^2N_c^2}{2}\frac{e^{-\frac{5}{4}\phi_h}}{\pi^3 u_h^3}
\end{split}\ee %
where $\phi_h$ is the value of the Dilaton field at the horizon and $N_c$ denotes the number of color in the gauge theory.
In this solution non-zero anisotropy parameter in the field theory is introduced by an axion field in the background corresponding to a
position-dependent $\theta$-term or, more precisely, $\theta=a z$. The $\theta$-term breaks the original isotropy of the system and forces the system into an anisotropic equilibrium state. It also
leads to a non-zero conformal anomaly meaning that the trace of the energy-momentum tensor is no longer zero and is proportional to the anisotropy parameter $a$ which has dimension of energy. In other words, in the field theory corresponding to the geometry \eqref{metrica}, the pressure in the $z$ and $\vec{x}$ directions are not equal. For more details see \cite{Mateos:2011tv}.
After finding the solution \eqref{metrica} numerically, our main task in the following is to understand how anisotropy parameter modifies the entanglement entropies $S_A^\bot$ and $S_A^\parallel$.
\subsection{Perpendicular case}
\begin{itemize}
\item \textbf{Fixed temperature:} In figure \ref{TfixoaT}(left), we have plotted $\Delta S^\bot$ in terms of $a/T$ for three different values of temperature. Evidently, $\Delta S^\bot>0$ meaning that $S_A^\bot>S_A^0$. Furthermore, the larger anisotropy parameter, the greater deviation.
We also observe that our numerical outcomes, the blue, red and black points in figure 2(left), are fitted with
\be\label{eoaTa}
\Delta S_A^\bot(a,T) = {K_1(T)}{(\frac{a}{T})^2}-K_2(T)(\frac{a}{T})^4,
\ee
\begin{table}[ht]
\caption{Coefficient of \eqref{eoaTa} and \eqref{eq2}}
\vspace{1 mm}
\centering
\begin{tabular}{c c c c c c c c}
\hline\hline
$T$ & $K_1$ & $K_2$ & $k_1$ & $k_2$\\
\hline\hline
0.4 & 0.024 & 3.27$\times 10^{-7}$ & 0.150 & 1.145$\times 10^{-5}$ \\
1.5 & 0.337 & 5.63$\times 10^{-5}$ & 0.150 &1.112$\times 10^{-5}$ \\
3.1 & 1.445 &9.42$\times 10^{-4}$ & 0.152 & 1.115$\times 10^{-5}$ \\
\hline
\end{tabular}\\[1ex]
\label{list}
\end{table}
where the values of functions $K_1(T)$ and $K_2(T)$ are listed in table \ref{list}.
Another valuable result comes out when one plots $\Delta S^\bot$ in terms of anisotropy parameter $a$ instead of $a/T$, see figure \ref{TfixoaT}(right). Surprisingly, it shows that the value of $\Delta S^\bot$, for various values of temperature, depends only on the anisotropy parameter and can be fitted with
\be\label{eq2} %
\Delta S^\bot(a)=k_1 a^2-k_2 a^4,
\ee %
where $k_1$ and $k_2$ are positive constant and their values are also presented in table \ref{list}. In fact, \eqref{eq2} indicates that, to a good approximation, $\Delta S^\bot$ is independent of temperature and the coefficients $k_1$ and $k_2$ are enough to compute the value of $\Delta S^\bot$ for given anisotropy parameter. However, the temperature dependence explicitly appears in \eqref{eoaTa} and
therefore these two equations, that is \eqref{eoaTa} and \eqref{eq2}, enforce us to consider $K_1(T)=k_1 T^2$ and $K_2(T)=k_2 T^4$.
Before closing this part a few comments are in case.
\begin{enumerate}
\item Although the values of $k_1$ or $k_2$ are not exactly the same for three values of temperature, one needs much greater precision in the numerical calculation to approach closer to the same value. In other words, the differences among the values of $k_1$ or $k_2$ would be more negligible with greater precision.
\item The value of $K_2$ is always too small compared to $K_1$. Thus the second term in \eqref{eoaTa}, or equivalently in \eqref{eq2}, does not change the value of the $\Delta S^\bot$ substantially for small values of anisotropy parameter. In other words, if one ignores this term, the maximum relative error, i.e. $\mid \frac{\Delta S^\bot-(\Delta S^\bot)_{k_2=0}}{\Delta S^\bot}\mid$, is about 0.23 for $a=50$.
\item Since the trace of the energy-momentum tensor is proportional to anisotropy parameter \cite{Mateos:2011tv}, one can conclude that the entanglement entropy is larger when the conformal symmetry is highly broken.
\item From \eqref{a1} and \eqref{eq2}, it turns out
\be %
S_A^\bot(a,T)=\frac{V_{d-1}}{G_N^{d+2}}(k_1 a^2-k_2 a^4)+S_A^0(T).
\ee
The first (second) term in the above equation is only a function of anisotropy parameter (temperature). Thus our results propose that the entanglement entropy in the anisotropic background we consider in this paper can be decomposed as
\be\label{temp} %
S_A^\bot(a,T)=\hat{S}_A(a)+S^0_A(T),
\ee
where
\be\label{final} %
\hat{S}_A(a)=\frac{V_{d-1}}{G_N^{d+2}}(k_1 a^2-k_2 a^4).
\ee %
In fact the anisotropy parameter and temperature do not communicate together.
\item The last point, perhaps the more important one, is that all above numerical outcomes are reliable in the range of $4T<a<50$.
\end{enumerate}
\begin{figure}[ht]
\begin{center}
\includegraphics[width=70mm]{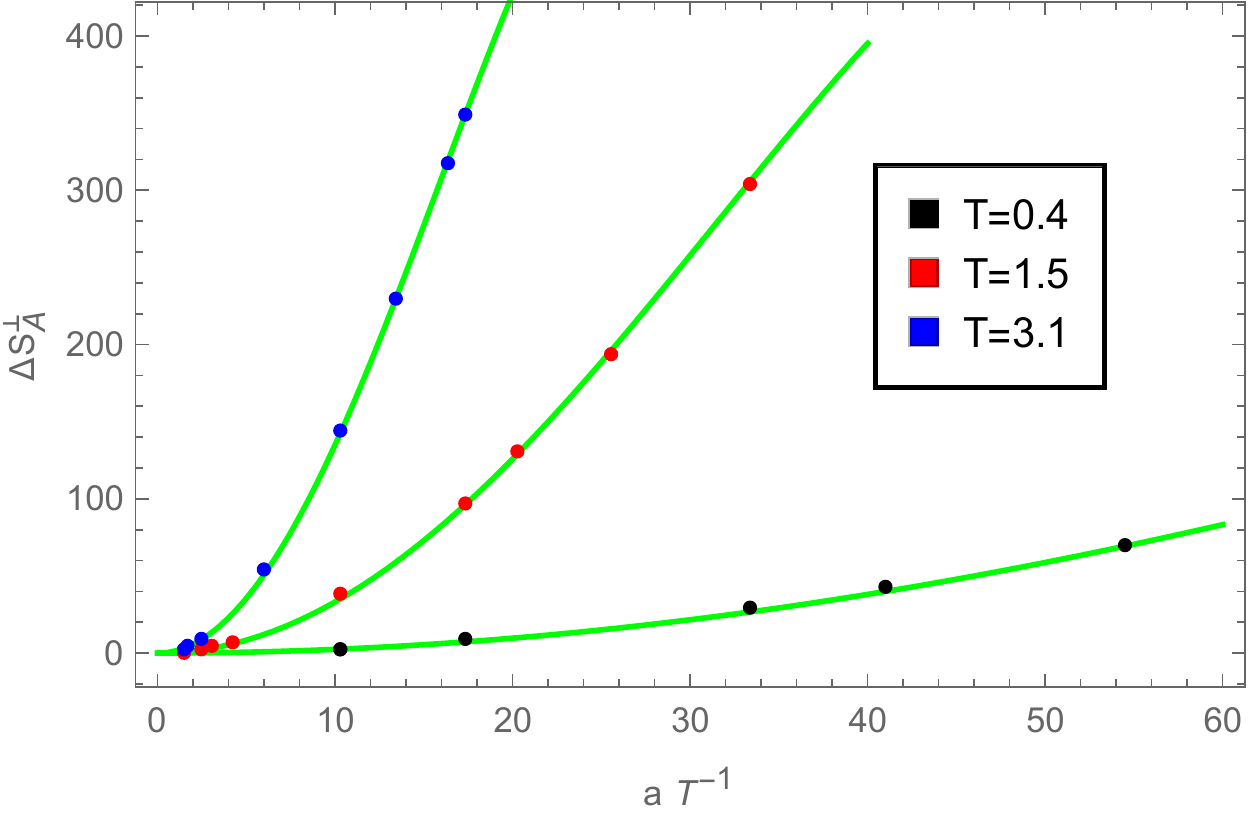}
\hspace{10 mm}
\includegraphics[width=70mm]{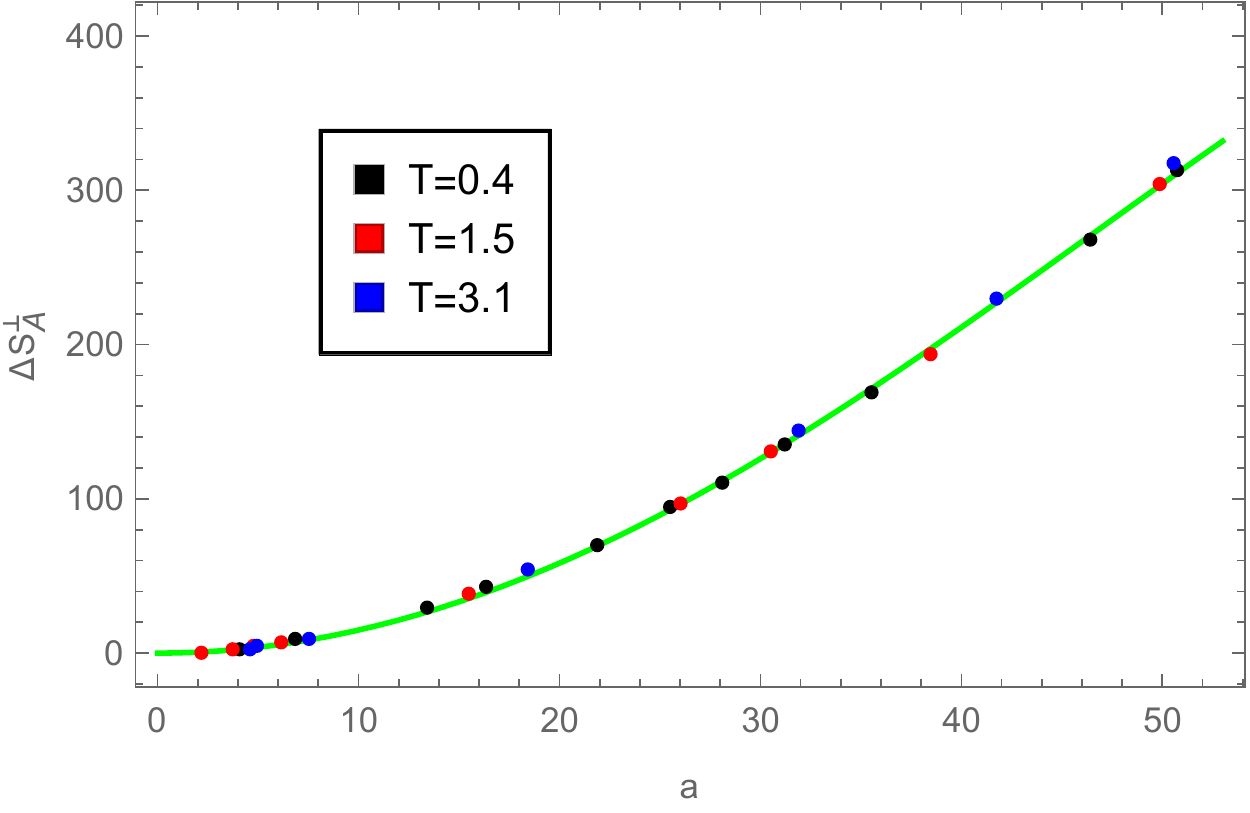}
\caption{$\Delta S_A^\bot$ in terms of $\frac{a}{T}$(left) and $a$(right) at fixed temperatures for $l=0.1$. The green curves show the fitted functions \eqref{eoaTa} and \eqref{eq2} in the left and right figure, respectively.}\label{TfixoaT}
\end{center}
\end{figure}
\item \textbf{Fixed entropy density:} Instead of temperature, the entropy density can be kept fixed, too. In order to have a dimensionless parameter we introduce $a s^{-1/3}$ where $s\equiv s/N_c^2$. We then plot the entanglement entropy in terms of this new dimensionless quantity and the results are shown in figures \ref{deltas,a,over,s}. Similar to the case of fixed temperature, we observe that the numerical data is fitted with
\be\label{ees} %
S_A^\bot(a,s)=\hat{K}_1(s)(as^{-1/3})^2-\hat{K}_2(s)(as^{-1/3})^4,
\ee %
where the values of $\hat{K}_1$ and $\hat{K}_2$ are listed in table \ref{list1}. Interestingly, although $\hat{K_1}$ and $\hat{K_2}$ are not equal to $K_1$ and $K_2$, numerical results in the figure \ref{deltas,a,over,s}(right) reveal that the entanglement entropy still satisfies in \eqref{eq2} for different values of fixed entropy density \textit{with unchanged coefficients} $k_1$ and $k_2$! Notice that in this case the temperature appears in \eqref{temp} can be found in terms of anisotropy parameter and entropy density by using \eqref{entro}. As a matter of fact, for given values of entropy density and anisotropy parameter, one can find the corresponding temperature. In short, we find out that the entanglement entropy, given by \eqref{eq2} or equivalently \eqref{temp}, is \textit{independent of the values of temperature and entropy density} in the desired range of anisotropy parameter. Similar comments argued in the case of fixed temperature can be also discussed in the current case and we do not repeat them here.
\end{itemize}
\begin{table}[ht]
\caption{Coefficient of \eqref{eq2} and \eqref{ees}}
\vspace{1 mm}
\centering
\begin{tabular}{c c c c c c c c}
\hline\hline
$s$ & $\hat{K}_1$ & $\hat{K}_2$ & $k_1$ & $k_2$\\
\hline\hline
1 &0.150 & 1.090$\times 10^{-5}$ & 0.150 & 1.090 $\times 10^{-5}$ \\
3 &0.310 & 4.526$\times 10^{-5}$ & 0.149 & 1.046 $\times 10^{-5}$ \\
5 &0.434 & 4.526$\times 10^{-5}$ & 0.148 & 1.013 $\times 10^{-5}$ \\
\hline
\hline
\end{tabular}\\[1ex]
\label{list1}
\end{table}
\begin{figure}[!ht]
\begin{center}
\includegraphics[width=70mm]{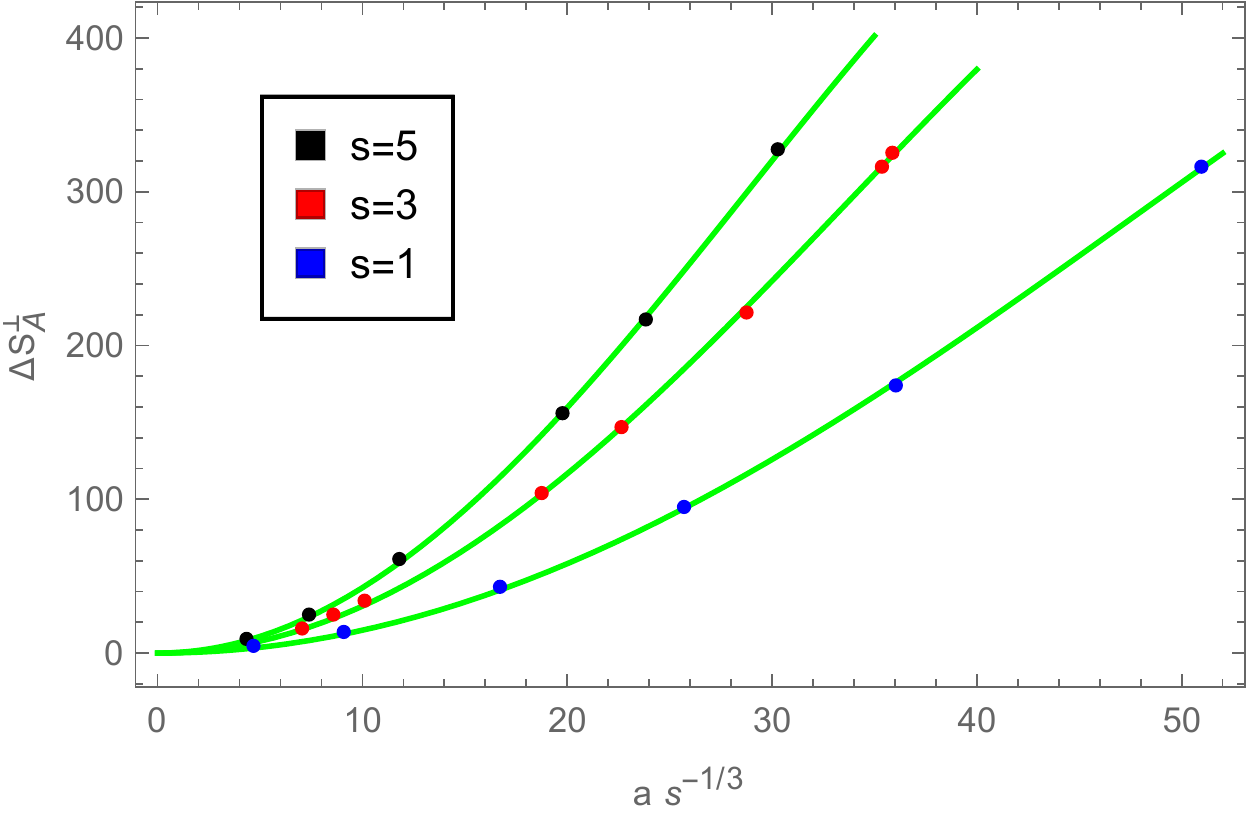}
\hspace{10 mm}
\includegraphics[width=70mm]{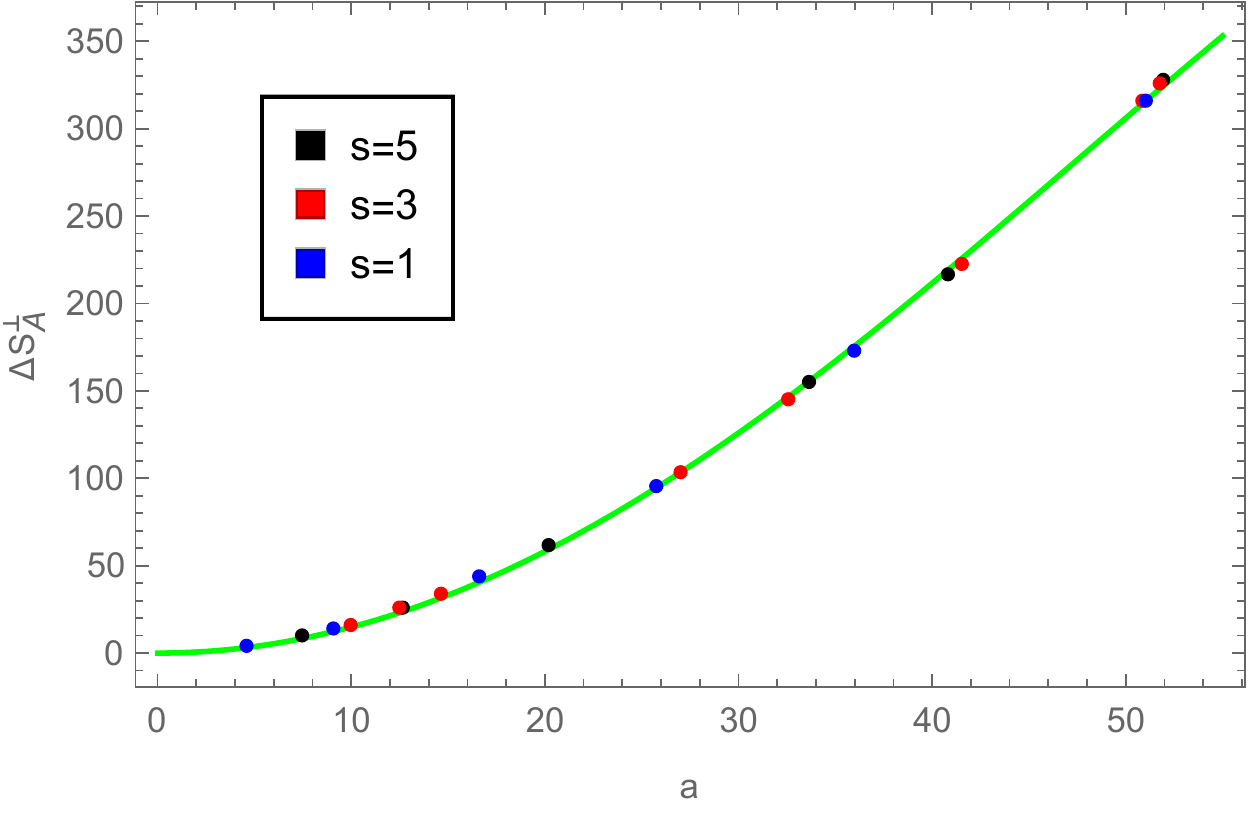}
\caption{$\Delta S_A^{\bot}$ in terms of $as^{-1/3}$(left) and $a$(right) at fixed entropy density for $l=0.1$. The green curves show the fitted functions \eqref{ees}(left) and \eqref{eq2}(right).}\label{deltas,a,over,s}
\end{center}
\end{figure}
\begin{figure}[!ht]
\begin{center}
\includegraphics[width=70mm]{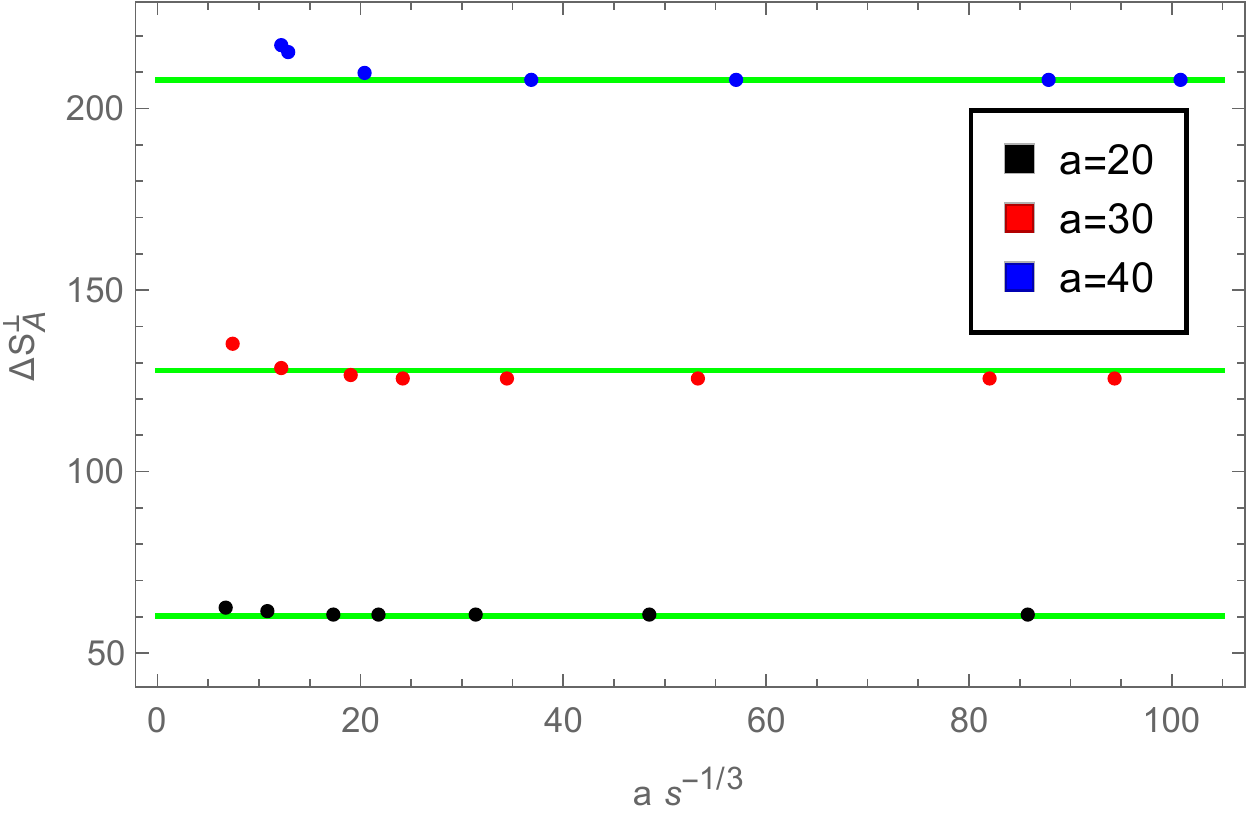}
\hspace{10 mm}
\includegraphics[width=70mm]{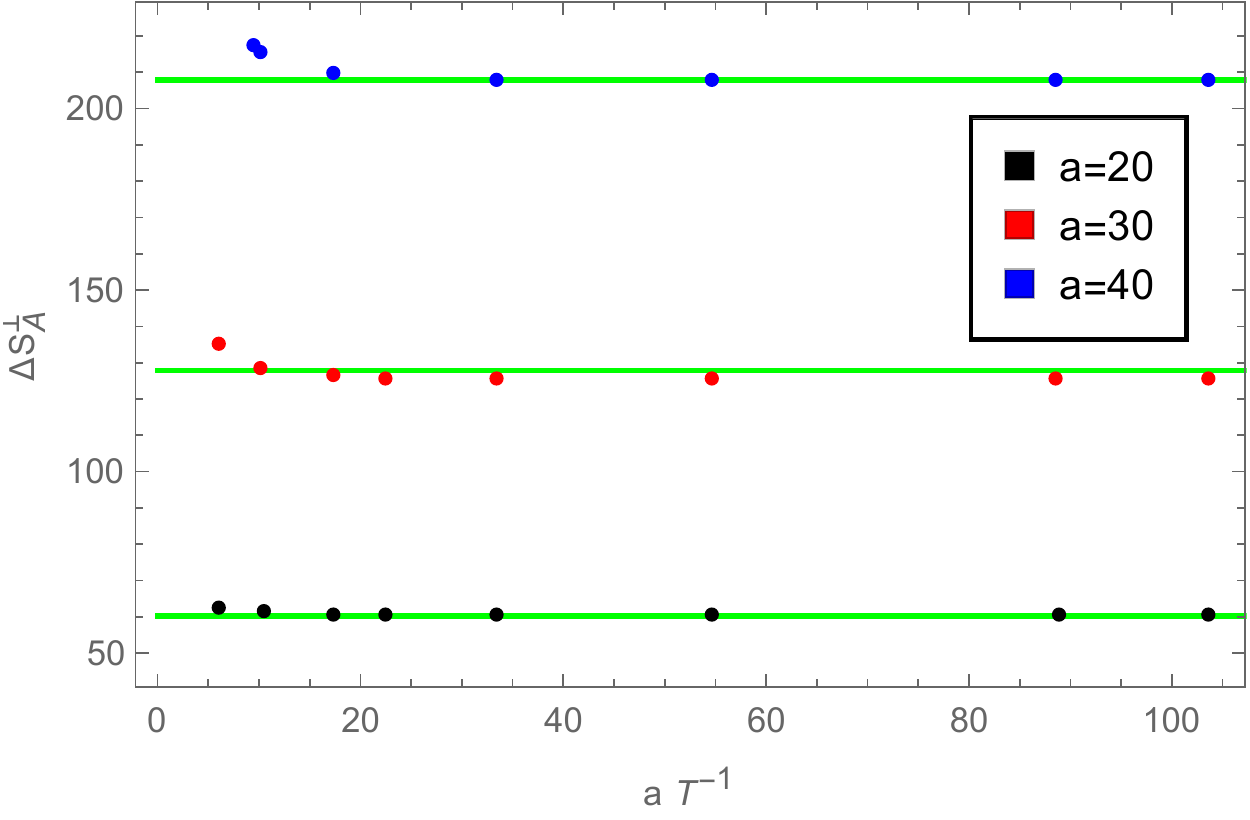}
\caption{$\Delta S^{\bot}_A$ as a function of $a s^{-1/3}$(right) and $\frac{a}{T} $(left) for $ l=0.1 $. 
}\label{a-fix-T-iso}
\end{center}
\end{figure}
Up to now, according to our numerical results we claim that $\Delta S_A^\bot$ is only a function of anisotropy parameter in the background we consider. As a crosscheck, in figure \ref{a-fix-T-iso} the entanglement entropy is plotted in terms of temperature and entropy density for fixed values of anisotropy parameters. As it is clearly seen, for large enough values of anisotropy parameter, that is $a>2s^{1/3}$ or $a>2T$, the entanglement entropy is given by \eqref{eq2} to a good approximation. In fact these figures approve our previous results and claim.
\subsection{Parallel case}
In this subsection we consider the case for which the entangling length is parallel to the anisotropy direction $z$. Figure \eqref{Tfixp-a} shows our results. The main difference between the parallel and perpendicular case is that in the parallel case $\Delta S^\parallel<0$ meaning that $S^\parallel_A<S_A^0$, opposite to the perpendicular case for which $S_A^\bot>S_A^0$. A few of significant outcomes can be summarized as follows:
\begin{figure}[!ht]
\begin{center}
\includegraphics[width=70mm]{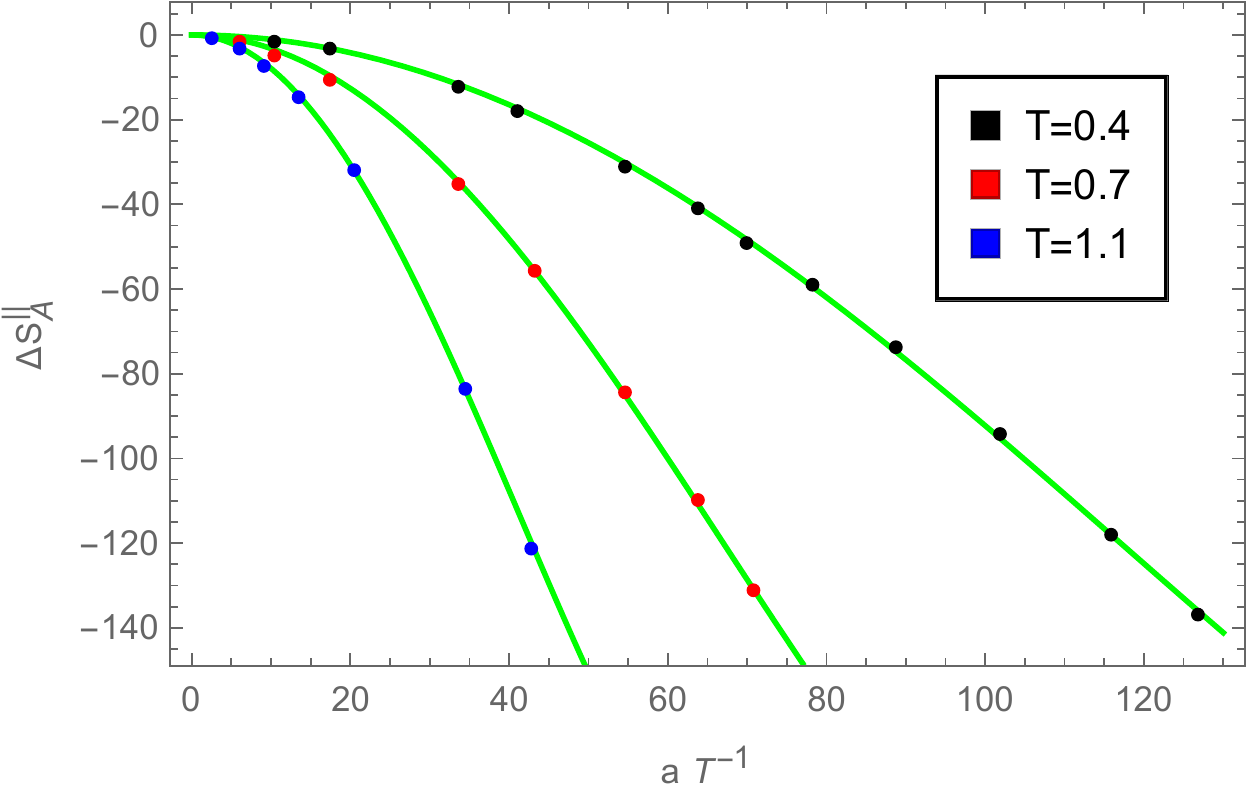}
\hspace{10 mm}
\includegraphics[width=70mm]{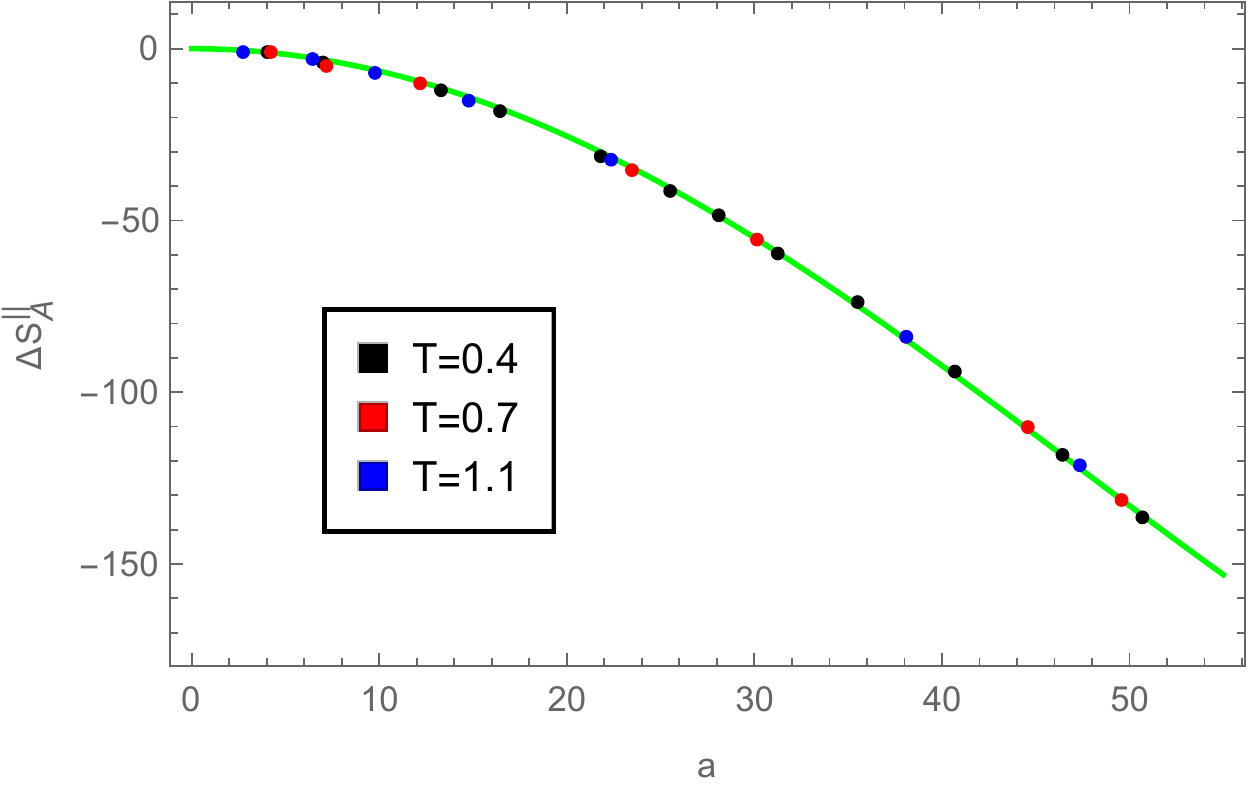}
\caption{$\Delta S^\parallel_A$ in terms of $\frac{a}{T}$(left) and $a$(right) at fixed temperatures for $l=0.1$. The green curves show the fitted functions.}\label{Tfixp-a}
\end{center}
\end{figure}
\begin{itemize}
\item In the desired range of anisotropy parameter the value of the parallel entanglement entropy $S_A^\parallel$ is always smaller than $S_A^0$ in both cases, when the temperature or entropy density is kept fixed.
\item For larger anisotropy parameter the deviation of $S_A^\parallel$ from the isotropic entanglement entropy becomes greater.
\item The entanglement entropy in the parallel case can be decomposed as
\be %
S_A^\parallel(a,T)=\tilde{S}_A(a)+S^0_A(T)
\ee
where $\tilde{S}_A(a)=\frac{V_{d-1}}{G_N^{d+2}}(-n_1 a^2+n_2 a^4)$, $n_1=0.065$ and $n_2=5\times10^{-6}$ for fixed value of temperature or entropy density cases.
\end{itemize}
In this section, our most important outcome is that for large enough $a/T$ (or $as^{-1/3}$) the entanglement entropy in the parallel and perpendicular direction can be decompose to a temperature dependent term, which is the entanglement entropy of the isotropic field theory at finite temperature, and an anisotropic part where is independent of temperature (or entropy density) to a good approximation. As we will see in the next section, this behavior persists in the high temperature limit, too.
\section{High Temperature Limit}
\begin{figure}
\begin{center}
\includegraphics[width=70mm]{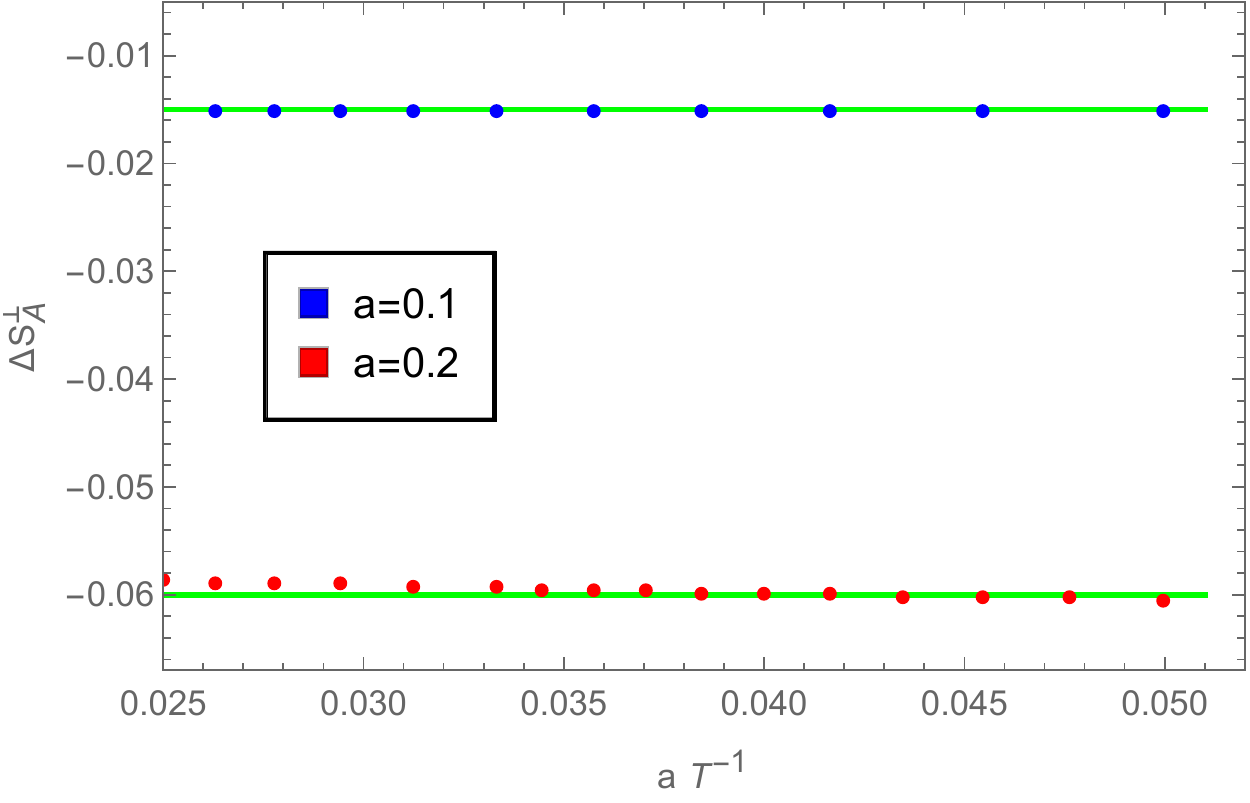}
\caption{$\Delta S^\bot_A$, introduced in (\ref{ah1}), in terms of $a/T$ for $l=0.1$.}\label{secondterm}
\end{center}
\end{figure}

The analytical solution for \eqref{metrica} has been introduced in \cite{Mateos:2011tv} in the high temperature limit, that is $T\gg a$. The metrics components are given by
\be\begin{split}\label{high}
\FF&=1-\frac{u^4}{u_h^4}+a^2\hat{\FF}_2(u), \cr
\B&=1+a^2\hat{\B}_2(u),\cr
\phi&=a^2\hat{\phi}_2(u),
\end{split}\ee
where
\be\begin{split}
\hat{\FF}_2(u)&=\frac{1}{24u_h^2}\bigg(8u^2(u_h^2-u^2)-10u^4\log2+(3u_h^4+7u^4)\log(1+\frac{u^2}{u_h^2})\bigg),\cr
\hat{\B}_2(u)&=-\frac{u_h^2}{24}\left(\frac{10u^2}{u_h^2+u^2}+\log(1+\frac{u^2}{u_h^2})\right),\cr
\hat{\phi}_2(u)&=-\frac{u_h^2}{4}\log(1+\frac{u^2}{u_h^2}).
\end{split}\ee
The temperature and entropy density is then obtained as
\be\begin{split} %
T&=\frac{1}{\pi u_h}+\frac{(5\log2-2)u_h}{48\pi^2}a^2, \cr
s&=\frac{\pi^2 N_c^2 T^3}{2}+\frac{N_c^2 T}{16}a^2.
\end{split}\ee
Using \eqref{eev} or \eqref{eep}, \eqref{metrica} and \eqref{high} and after simple calculation, it turns out
%
%
%
%
\bse
\begin{align}
\label{ah2} {S^ \bot_{A} }&=\frac{V_{d - 1}}{2 G_N^{d+2}}\int_0^{u_*}\frac{(1-\frac{5}{2}a^2\hat{\phi}_2(u))(1-\h \frac{a^2\hat{\FF}_2(u)}{1-\frac{u^4}{u_h^4}})(1+\frac{5}{4} a^2\frac{\hat{\phi}_2(u) u^6_*-\hat{\phi}_2(u_*)u^6}{u^6_*-u^6})}{\sqrt{1-\frac{u^4}{u_h^4}}\sqrt{u^{-6} - u_*^{-6 }}} du,\\
\label{ah1} &=S_A^0+\frac{V_{d - 1}}{2 G_N^{d+2}}a^2\int_0^{\hat{u}_*}\frac{-\frac{5}{2}\hat{\phi}_2(u)-\h \frac{\hat{\FF}_2(u)}{1-\frac{u^4}{\hat{u}_h^4}}+\frac{5}{4} \frac{\hat{\phi}_2(u) \hat{u}^6_*-\hat{\phi}_2(\hat{u}_*)u^6}{\hat{u}^6_*-u^6}}{\sqrt{1-\frac{u^4}{\hat{u}_h^4}}\sqrt{u^{-6} - \hat{u}_*^{-6 }}} du,
\end{align}
\ese
or in the parallel case
\begin{equation}
\begin{split}\label{sparallel}
{S^ \parallel_{A} }
&=S_{A}^0+\frac{V_{d - 1}}{2 G_N^{d+2}}a^2\int_0^{\hat{u}_*}\frac{-2\hat{\phi}_2(u)-\h \frac{\hat{\FF}_2(u)}{1-\frac{u^4}{\hat{u}_h^4}}+\frac{5}{4} \frac{\hat{\phi}_2(u) \hat{u}^6_*-\hat{\phi}_2(\hat{u}_*)u^6}{\hat{u}^6_*-u^6}}{\sqrt{1-\frac{u^4}{\hat{u}_h^4}}\sqrt{u^{-6} - \hat{u}_*^{-6 }}} du.
\end{split}
\end{equation}
It is important to notify that $u_*$ in \eqref{ah2} denotes the turning point of the extremal surface $\gamma_A$ (see figure \ref{fff}) in the anisotropic background which is not clearly equal to the corresponding one in the isotropic case, that is $\hat{u}_*$. However, since our results indicate that their difference, i.e. $\max(\frac{|u_*-\hat{u}_*|}{\hat{u}_*})\sim 10^{-4}$, is so small, it is a good approximation to neglect this difference and we therefore replace $u_*$ by $\hat{u}_*$ in \eqref{ah1}. The same argument is also valid for $u_h$ and $\hat{u}_h$ corresponding to horizon radius in anisotropic and isotropic background. Then the term which is independent of $a$ in \eqref{ah2} reduces to the entanglement entropy in the isotropic filed theory, that is the first term in \eqref{ah1}. As a result, the first term in \eqref{ah1} is independent of anisotropy parameter and is equal to the entanglement entropy in the isotropic field theory. Although the second term in \eqref{ah1} explicitly depends on the $u_h$ (or equivalently $\hat{u}_h$), this term, which is equal to $\Delta S_A^\bot$ according to \eqref{a1}, is almost independent of temperature as it is supported by figure \ref{secondterm}.
We can continue the same discussion for the parallel case to find \eqref{sparallel}. Briefly, in high temperature limit, the entanglement entropy for both parallel and perpendicular cases decomposes into two parts: the first (second) part is (almost) independent of anisotropy parameter (temperature) in agreement with our previous results.
\begin{figure}
\begin{center}
\includegraphics[width=70mm]{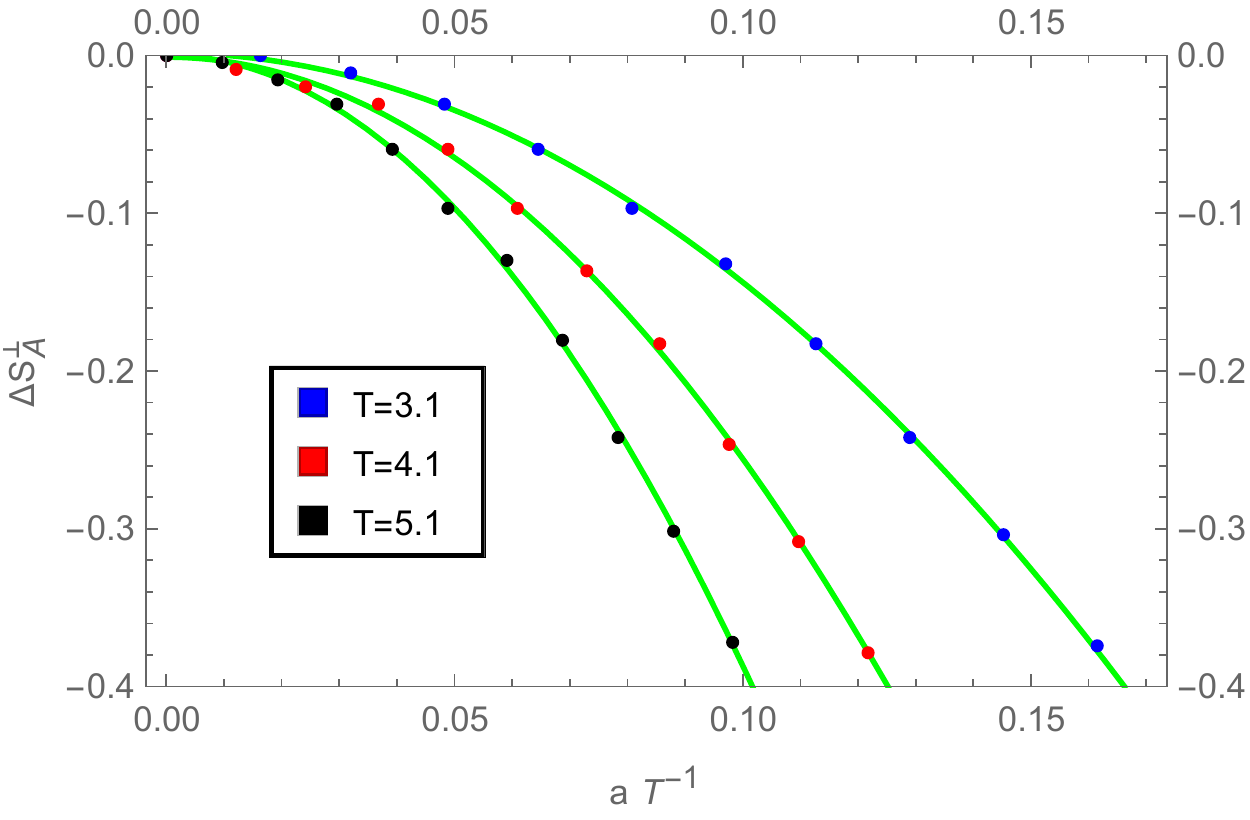}
\hspace{10 mm}
\includegraphics[width=70mm]{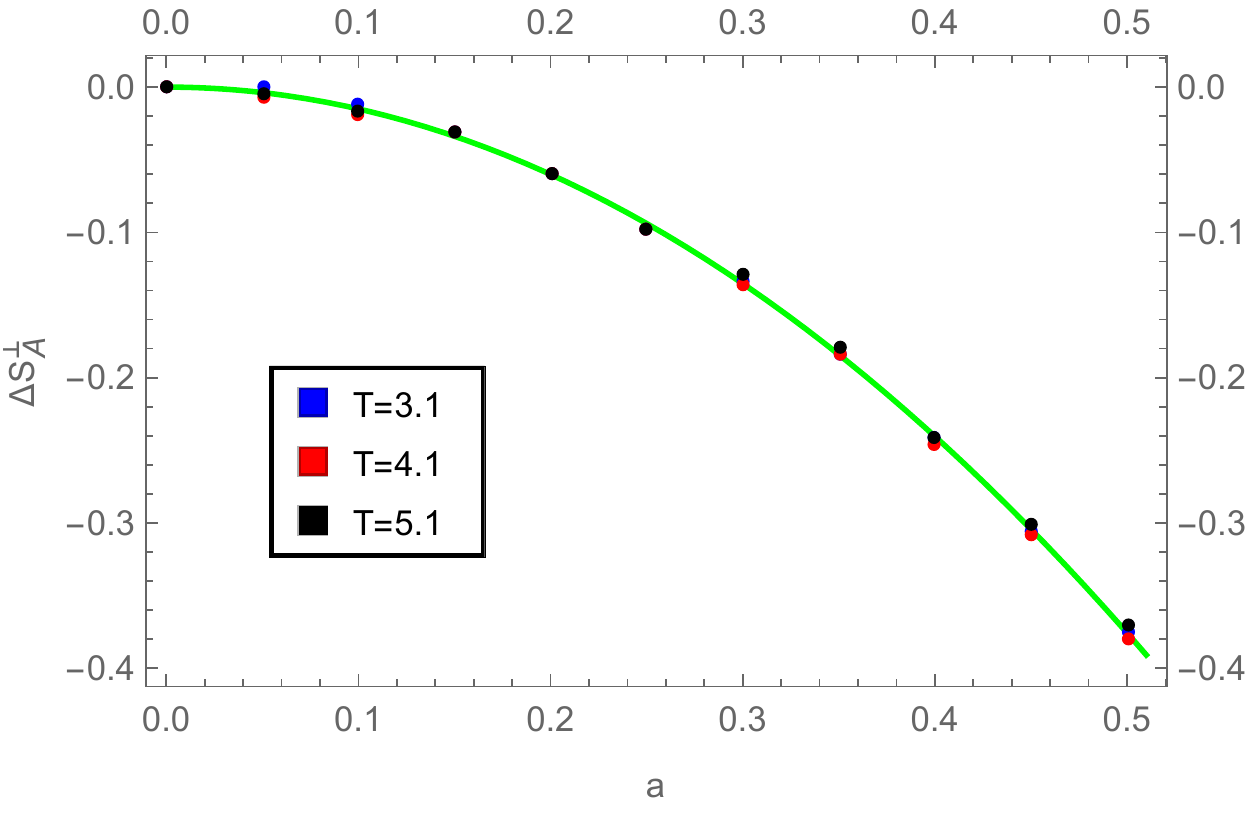}
\caption{$\Delta S^{\bot}_A$ as a function of $a T^{-1}$ (right) and $a $ (left) in high temperature limit for $ l=0.1 $. The green curves show fitted functions. 
}\label{hightemp}
\end{center}
\end{figure}
In contrast to the general case, in high temperature limit figure \eqref{hightemp} shows that $S_A^\bot<S_A^0$. Unfortunately, we cannot find the entanglement entropy between two regions we have discussed here and in the previous section. However, our results indicate that at fixed temperature the entanglement must change sign at a specified value of anisotropy parameter. Similar behavior has been also observed in \cite{Rahimi:2016bbv}.
For the parallel case, opposite to the perpendicular case, the entanglement entropy is \textit{always} smaller than isotropic one. In high temperature limit, the difference between $\Delta S_A^\bot$ and $\Delta S_A^\parallel$ is negligible and therefore resulted figure for parallel case is similar to the figure \eqref{hightemp} and we do not repeat them here.
\section*{Acknowledgement}
We would like to thank School of Physics of Institute for research in fundamental sciences (IPM) for the research facilities and environment.


\begin{thebibliography}{99}
\bibitem{Maldacena}
J.~M.~Maldacena,
``The large N limit of superconformal field theories and supergravity,''
Adv.\ Theor.\ Math.\ Phys.\ {\bf 2} (1998) 231
[Int.\ J.\ Theor.\ Phys.\ {\bf 38} (1999) 1113][arXiv:hep-th/9711200];
S.~S.~Gubser, I.~R.~Klebanov and A.~M.~Polyakov,
``Gauge theory correlators from non-critical string theory,''
Phys.\ Lett.\ B {\bf 428} (1998) 105[arXiv:hep-th/9802109];
E.~Witten,
``Anti-de Sitter space and holography,''
Adv.\ Theor.\ Math.\ Phys.\ {\bf 2} (1998) 253[arXiv:hep-th/9802150].
\bibitem{solana}
J.~Casalderrey-Solana, H.~Liu, D.~Mateos, K.~Rajagopal and U.~A.~Wiedemann,
``Gauge/String Duality, Hot QCD and Heavy Ion Collisions,''
book:Gauge/String Duality, Hot QCD and Heavy Ion Collisions. Cambridge, UK: Cambridge University Press, 2014
[arXiv:1101.0618 [hep-th]].
\bibitem{Ryu:2006ef}
S.~Ryu and T.~Takayanagi,
``Aspects of Holographic Entanglement Entropy,''
JHEP {\bf 0608}, 045 (2006)
[hep-th/0605073];
\bibitem{Mateos:2011tv}
D.~Mateos and D.~Trancanelli,
``Thermodynamics and Instabilities of a Strongly Coupled Anisotropic Plasma,''
JHEP {\bf 1107}, 054 (2011)
[arXiv:1106.1637 [hep-th]].
\bibitem{thesis}
M.~Jonker,
"Entanglement entropy of coupled harmonic oscillators,"
studenttheses.library.uu.nl › Library › Master's / Bachelor's Theses Online
\bibitem{Headrick:2007km}
M.~Headrick and T.~Takayanagi,
``A Holographic proof of the strong subadditivity of entanglement entropy,''
Phys.\ Rev.\ D {\bf 76}, 106013 (2007)
[arXiv:0704.3719 [hep-th]];
M.~Headrick,
``Entanglement Renyi entropies in holographic theories,''
Phys.\ Rev.\ D {\bf 82}, 126010 (2010)
[arXiv:1006.0047 [hep-th]];
M.~R.~Mohammadi Mozaffar and A.~Mollabashi,
``Entanglement in Lifshitz-type Quantum Field Theories,''
JHEP {\bf 1707}, 120 (2017)
[arXiv:1705.00483 [hep-th]];
M.~R.~Mohammadi Mozaffar, A.~Mollabashi, M.~M.~Sheikh-Jabbari and M.~H.~Vahidinia,
``Holographic Entanglement Entropy, Field Redefinition Invariance and Higher Derivative Gravity Theories,''
Phys.\ Rev.\ D {\bf 94}, no. 4, 046002 (2016)
[arXiv:1603.05713 [hep-th]];
M.~R.~Mohammadi Mozaffar, A.~Mollabashi and F.~Omidi,
``Holographic Mutual Information for Singular Surfaces,''
JHEP {\bf 1512}, 082 (2015)
[arXiv:1511.00244 [hep-th]].
\bibitem{Ali-Akbari:2014xea}
M.~Ali-Akbari and D.~Allahbakhshi,
``Meson Life Time in the Anisotropic Quark-Gluon Plasma,''
JHEP {\bf 1406}, 115 (2014)
[arXiv:1404.5790 [hep-th]];
M.~Ali-Akbari and S.~F.~Taghavi,
JHEP {\bf 1504}, 181 (2015)
[arXiv:1408.6361 [hep-th]];
M.~Ali-Akbari and H.~Ebrahim,
``Chiral symmetry breaking: To probe anisotropy and magnetic field in quark-gluon plasma,''
Phys.\ Rev.\ D {\bf 89}, no. 6, 065029 (2014)
[arXiv:1309.4715 [hep-th]];
K.~Bitaghsir Fadafan and R.~Morad,
``Jets in a strongly coupled anisotropic plasma,''
Eur.\ Phys.\ J.\ C {\bf 78}, no. 1, 16 (2018)
[arXiv:1710.06417 [hep-th]];
K.~Bitaghsir Fadafan, D.~Giataganas and H.~Soltanpanahi,
``The Imaginary Part of the Static Potential in Strongly Coupled Anisotropic Plasma,''
JHEP {\bf 1311}, 107 (2013)
[arXiv:1306.2929 [hep-th]];
M.~Chernicoff, D.~Fernandez, D.~Mateos and D.~Trancanelli,
``Drag force in a strongly coupled anisotropic plasma,''
JHEP {\bf 1208}, 100 (2012)
[arXiv:1202.3696 [hep-th]];
M.~Chernicoff, D.~Fernandez, D.~Mateos and D.~Trancanelli,
``Jet quenching in a strongly coupled anisotropic plasma,''
JHEP {\bf 1208}, 041 (2012)
[arXiv:1203.0561 [hep-th]];
S.~Chakraborty and N.~Haque,
``Holographic quark-antiquark potential in hot, anisotropic Yang-Mills plasma,''
Nucl.\ Phys.\ B {\bf 874}, 821 (2013)
[arXiv:1212.2769 [hep-th]].
\bibitem{Rahimi:2016bbv}
M.~Rahimi, M.~Ali-Akbari and M.~Lezgi,
``Entanglement entropy in a non-conformal background,''
Phys.\ Lett.\ B {\bf 771}, 583 (2017)
[arXiv:1610.01835 [hep-th]].
\end{thebibliography}
\end{document}